\documentclass[]{pasj02} 

\usepackage[switch,mathlines]{lineno} 
\usepackage{url}
\usepackage{graphicx}
\let\degree\relax
\usepackage{gensymb}

\jyear{2025}
\Received{}
\Accepted{}

\begin{document} 
\nolinenumbers

\title{Detection of the 2021 Arid Meteor Shower on Maunakea, Hawai'i}

\author{Ichi
\textsc{Tanaka},\altaffilmark{1}\altemailmark\orcid{0000-0002-4937-4738} \email{ichi@naoj.org} 
 Hitoshi \textsc{Hasegawa},\altaffilmark{2}
 Toyokazu \textsc{Uda},\altaffilmark{3}
 Mikiya \textsc{Sato},\altaffilmark{4} 
 Jun-ichi \textsc{Watanabe}\altaffilmark{4}
 and
 Masanobu \textsc{Higashiyama},\altaffilmark{5}
}
\altaffiltext{1}{Subaru Telescope, National Astronomical Observatory of Japan, 650 North A'ohoku Place, Hilo, Hawai'i, 96720, U.S.A.}
\altaffiltext{2}{Suginami-ku, Tokyo 168-0074, JAPAN}
\altaffiltext{3}{Aiharasoft, Co., Sagamihara, Kanagawa 252-0141, Japan}
\altaffiltext{4}{National Astronomical Observatory of Japan, 2-21-1 Osawa, Mitaka, Tokyo 181-8588, Japan}
\altaffiltext{5}{The Asahi Shimbun Company, 5-3-2 Tsukiji, Chuo-ku, Tokyo, Japan}


\KeyWords{methods: observational Planetary Systems -- meteors:individual (ARD, \#1130)}  

\maketitle

\begin{abstract}
We report the successful detection of the "Arid" Meteor Shower (IAU\#1130 ARD), predicted to emerge for the first time in 2021, using a publicly accessible YouTube live camera developed by us. This live camera, installed on the Subaru Telescope dome in the summit region of Maunakea, Hawai'i, features a wide field of view ($70^{\circ} \times 40^{\circ}$) and high sensitivity, capable of observing stars fainter than 6th magnitude. Meteor detection was performed in two ways: visual inspection by citizen viewers and subsequent validation through automated detection. As a result, we confirmed that the number of meteors appearing from near the predicted radiant increased by more than six times ($\sim9\sigma$) compared to the preceding and following days. Our observation time was 4-5 hours after the predicted peak (solar longitude = 193.9$^{\circ}$), providing clear data indicating that the activity had not yet declined. Optical observations at this time from the Northern Hemisphere are extremely limited and unique, making our observation point valuable. The meteors are characterized as slow and faint appearance, but several brighter meteors with wakes were also observed. Simulations tracing the dust trails from the parent body, Comet 15P/Finlay, suggest that our detection can be explained by either the dust trails released in 2008 or 2014, both requiring high ejection velocities. However, during the comet's 2008 return, its activity was exceptionally quiet, making a high-velocity dust ejection questionable. On the other hand, multiple large outbursts were observed during the 2014 return, at which time a certain amount of high-velocity dust release is expected. We conclude that the dust source of the meteor shower detected in Hawai'i this time is likely attributable to high-velocity ($\sim$67~m s$^{-1}$) dust ejected during the 2014 outburst.
\end{abstract}


\section{Introduction}
The recent advancements in detector technology are remarkable, and astronomy has significantly benefited from them. This wave has also reached the field of meteor observation, where citizen scientists play a crucial role. Multiple global-scale meteor observation networks utilizing high-sensitivity cameras have been established in various locations (such as SONOTACO, CAMS, and GMS, \cite{2009JIMO...37...55S}; \cite{2011Icar..216...40J}; \cite{2013Icar..225..614W}; \cite{2021MNRAS.506.5046V}), revolutionizing meteor astronomy.

The "Subaru-Asahi StarCam" (hereafter the StarCam\footnote{\url{https://www.naoj.org/PIO/LiveCam/cam_redirect.html}}), installed at the Subaru Telescope site in the summit region of Maunakea, is a collaborative outreach project between NAOJ and the Asahi Shimbun, a Japanese newspaper company. Its aim is to deliver the unpolluted night sky to people worldwide by placing a cutting-edge, high-sensitivity camera at Maunakea, one of the world's premier astronomical observation sites. Soon after its operation began, we realized that the camera's characteristics -- wide field of view, high sensitivity, and real-time streaming -- could make significant contributions to meteor astronomy (Tanaka et al. 2025, hereafter Paper I). To further advance this verification, we attempted to detect the Arid meteor shower, which was predicted to appear for the first time in October 2021, by leveraging citizen power through the same platform. This paper describes this endeavor.

The parent body of the Arid meteor shower, Comet 15P/Finlay, is a Jupiter-family periodic comet with a period of six years. Its perihelion is at 0.976~AU, and its Minimum Orbit Intersection Distance ranks as the sixth smallest among known comets, exhibiting characteristics of a Near-Earth Object. For this reason, the existence of associated meteor showers has long been a subject of discussion. \citet{1999MNRAS.310..168B} traced the orbital evolution of the parent comet back to 1585 and investigated the possibility of past meteor showers by following the evolution of its dust trails. As a result, they found that dust particles ejected from the comet during the period when its perihelion was close to 1 AU (between 1886 and 1996) were swept outside Earth's orbit due to Jupiter's perturbation, attributing this as one of the reasons for the low meteor activity in the past. Furthermore, a re-analysis of radio survey databases conducted between 1960 and 1970 concluded that there was no indication of meteoroids originating from 15P/Finlay.

Comet 15P/Finlay had originally shown signs of gradually declining cometary activity, but it underwent a strong outburst during its 2014 return, increasing its brightness by over 100 times (e.g., \cite{2015ApJ...814...79Y}, \cite{2016AJ....152..169I}). Consequently, meteor showers resulting from the Earth's encounter with dust ejected during the 2014/15 outburst garnered renewed attention. Dust trail orbital evolution simulations by our Co-Investigator (hereafter Co-I), M.S., and others indicated the possibility that the 2014 dust trail would intersect Earth, making a meteor shower originating from Comet Finlay observable for the first time in history (\cite{2015ApJ...814...79Y}, and references therein). More recently, \citet{2020JIMO...48...29V} reported their more detailed simulations, forecasting meteor shower activity associated with the 1995 and 2014/2008 dust trails on September 29 and October 7 UT, 2021, respectively. Observational confirmation of the birth and scale of a new meteor shower, which had no prior activity history, in 2021 is crucial for validating dust trail dynamical evolution models.

The predicted radiant ($\alpha, \delta = 256, -48$) of this new meteor shower, located in the Constellation Ara, is observable from Hawai'i in the low southern sky during evening hours, at an altitude of 13 degrees or less. Although the predicted peak time was before sunset in Hawai'i, the astronomical twilight time was still 4 hours after the predicted peak. Therefore, there was a possibility that the activity could still be observed from Hawai'i if it remained active. We decided to attempt the observation of this new meteor shower as a science demonstration for this camera, collaborating with volunteer viewers. According to \citet{2020JIMO...48...29V}, because the impact velocity of the cometary dust with Earth is very low at 10.5~km s$^{-1}$, the expected brightness of the appearing meteors was very faint. This characteristic also makes it an excellent demonstration of our camera's high-sensitivity and wide-field observation capabilities.

The structure of this paper is as follows. Chapter 2 discusses the data. Chapter 3 describes the data evaluation methods. Two independent methods are adopted for evaluation: visual inspection of videos by volunteers and an automated detection method. Chapter 4 presents the results, and Chapter 5 discusses the result with the aid of the dust-trail simulation. Chapter 6 summarizes the discussion. Dates for results are described in UT, but HST (=UT-10) is also used when discussing individual observations.

\section{Observing System and the Data}\label{sec:2}

Details of the camera used for observation (Subaru-Asahi StarCam) are extensively described in Paper I, so only essential information is summarized here. This camera is installed on the catwalk of the Subaru Telescope, located in the summit area of Maunakea, Hawai'i, at an altitude of approximately 4150~m. The field of view is set almost directly to the east, covering 70$^{\circ} \times 40^{\circ}$. Table~\ref{tab:tbl1} summarises the basic site information. Its original purpose was to share the Maunakea night sky with the public and streams daily videos to YouTube as a color live camera at 30~fps (though effectively 1/15~s under low light conditions). With its high ISO sensitivity of 256000, it can detect 7th magnitude stars even with an exposure of 1/15~s, which is automatically set under low light ("Phase II" in Paper I; no Soft Filter is attached). 

As mentioned earlier, the primary mission of this camera is to share Maunakea's world-class night sky with the public. However, its wide field of view, high sensitivity, and high temporal resolution recognized its high potential for meteor observation. As stated in the Introduction, this Arid meteor shower observation campaign also serves as a demonstration of this camera's capabilities. Specifically, its ability to detect faint meteors is considered to match the characteristics of the Arid meteor shower.

\begin{table}
  \tbl{Camera Location.\footnotemark[$*$] }{%
  \begin{tabular}{cc}
      \hline
      Latitude & $19.8256~\degree$ \\ 
      Longitude &  $-155.4758~\degree$\\ 
      Altitude of Camera & $4153~m$ \\
      Field of View (FoV) & $40~\degree \times 70~\degree$ \\
      FoV Center (Az \& El) & +81.2~\degree, 19.6~\degree  \\
      Sky Fraction & 77\% \\
    \hline
    \end{tabular}}\label{tab:tbl1}
\begin{tabnote}
\footnotemark[$*$] WGS84  \\ 
\end{tabnote}
\end{table}

StarCam's observation area includes the low-altitude region, encompassing the Maunakea horizon. Generally, observation conditions at lower altitudes are poor due to decreased atmospheric transparency and light pollution. However, Maunakea is a world-class observation site with the negligible effects of light pollution. Furthermore, the observation site's altitude of 4,150~m dramatically reduces atmospheric absorption with a V-band absorption coefficient of only 0.11 (\cite{1987PASP...99..887K}).

The Arid meteor shower was expected to have many faint meteors, and indeed, many of the detected meteors were dim. It is worth mentioning that their detection would have been significantly more difficult without an excellent condition of the site.

We'll describe the weather condition over the three days of the campaign observation. The 6th and 7th (UT) were clear, but on the 8th, there were some thin cirrus clouds, and minor fraction of the sky were afffected by them. The new moon was on the 6th, so there was no lunar interference.

The altitude of the Arids radiant in Hawai'i was 13 degrees at 19:00 HST, when it was getting dark (twilight was at 18:53 PM, with a radiant altitude of 12 degrees). By 20:30 PM, the altitude was only 3 degrees, so we set the evaluation slot between 19:00 PM and 20:30 HST.

\section{Evaluation Method}\label{sec:3}
According to forecasts by our Co-I (MS) and \citet{2020JIMO...48...29V}, the Arid meteor shower had two activity windows on September 29 and October 7 (UT). \citet{2021eMetN...6..531J} reported that meteors were actually observed during the September 29 window\footnote{See also \url{http://www.cbat.eps.harvard.edu/iau/cbet/005000/CBET005046.txt}}. In response to this, an emergency volunteer-led meteor shower detection campaign was organized.

Data from the three days surrounding the predicted peak date of October 7 (UT) was recorded in 1080p HD at 30~fps, and visual counts were performed by volunteers. Subsequently, the same data was re-evaluated using the team's proprietary automatic detection software, and the final conclusion was reached by comparing both results.

\subsection{Eye-ball Evaluation}\label{ssec:3_1}
At the time the meteor shower was predicted, our software for automatically detecting meteors from YouTube live-streaming data had not yet been developed. The only method available for trajectory estimation was human visual inspection. We put out a call for volunteers among the viewers and received seven applications. We specifically sought volunteers not only to leverage the complementary effect of having multiple pairs of eyes, but also because this camera was intended for public outreach for the Maunakea Observatories, and we wanted to provide an opportunity for public participation in our scientific work.

The evaluation method was as follows. First, we defined an evaluation period of three days, including one day before and one day after the predicted peak date of the meteor shower, which was October 6 (HST) (= October 7 UT). The video recordings from October 5, 6, and 7 were each divided into three time slots: 19:00–19:30, 19:30–20:00, and 20:00–20:30 (HST).

The video recordings for each time slot were randomly assigned to the volunteers. Each volunteer was asked to watch all three days of the assigned time slot. Since there were seven evaluators, a specific time slot on a specific day was generally evaluated by two people, but the remaining volunteer was assigned a second time slot.

\begin{figure}
 \begin{center}
 \includegraphics[width=0.48\textwidth]{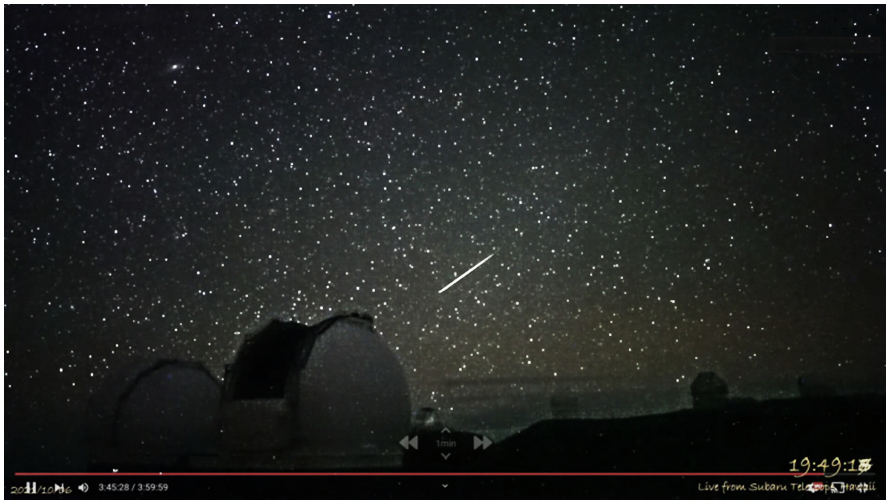}
\vspace{10mm} 
 \includegraphics[width=0.48\textwidth]{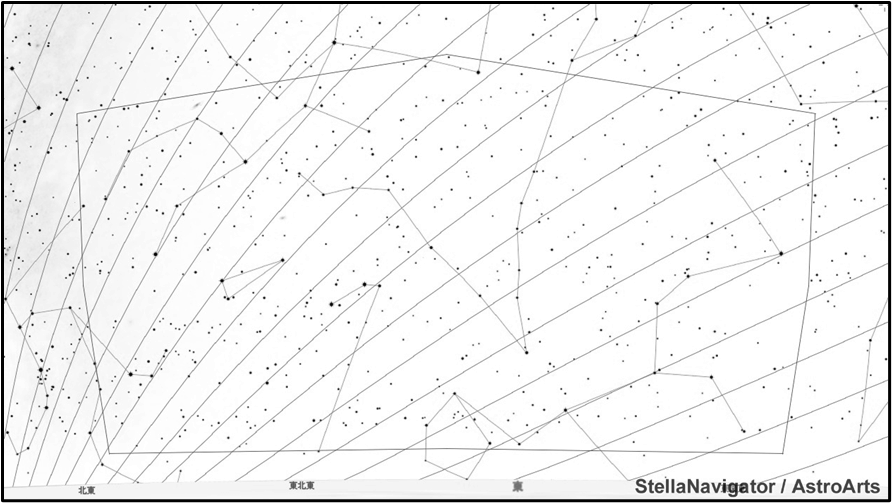}
 \end{center}
\caption{Evaluation of video data (Top) and radiant plot for trajectory determination (Bottom). Note that the bottom figure is expressed as negative images for better visibility. The trajectory diagram is created using the script function of AstroArts' Stellar Navigator Software\footnotemark. The zenith attraction effect is included in the calculation for this plot.
{Alt text: This figure explains how the volunteers evaluated each meteor if it is the Arids or not.} 
}\label{fig:fig1}
\end{figure}

We sent the recorded video URL addresses\footnote{They are available below: 10/5, \url{https://youtu.be/ssIgkJMwQLo}; 10/6, \url{https://youtu.be/QK5yH2o_J3k}, \url{https://youtu.be/u5N5Ogxo-ew}, \url{https://youtu.be/E7H_6IQZGsI}; 10/7, \url{https://youtu.be/pnwZDoBHETo}} to the evaluators and asked them to count potential Arid meteors and sporadic meteors for each 10-minute interval within their assigned time slots. To determine whether a meteor was a candidate for the shower, they were provided with a radiant plot created by one of our Co-Is (M.S.) which included the effects of zenith attraction (figure~\ref{fig:fig1}). For each candidate shower meteor, evaluators were asked to record the time of appearance to allow for later verification. After compiling these counts, we (I.T. and M.S.) performed a follow-up assessment to make a final shower determination.

\footnotetext{See \url{https://www.astroarts.co.jp/products/software.shtml}}

For the data aggregation, we adopted a method where we first took the maximum and minimum counts from multiple evaluators for each 10-minute interval within a given time slot. We then primarily used the maximum value. We chose this approach because we believed the main source of error was not misidentifying a different meteor as a shower meteor, but rather simply missing faint meteors with short trajectories, which were common. Any meteor that evaluators commented on as being difficult to judge was re-evaluated by us (I.T. and M.S.) after the initial compilation, and our findings were incorporated into the final results.

Table~\ref{tab:tbl2} summarizes the relationship between the volunteers and the time slots they evaluated. In the table, the bold font indicates volunteers who were specifically designated to a time slot, while the other individuals participated entirely on their own initiative. It is noteworthy to mention the evaluators' dedication. All volunteers completed the evaluation for their assigned time slots, and more than half of them also evaluated additional time slots beyond their assignments. Furthermore, two volunteers consistently evaluated all data throughout the entire observation period.

\begin{table}
  \tbl{Volunteers' Assignment for Evaluation }{%
  \begin{tabular}{lccc}
      \hline
     Time Slot (HST)\footnotemark[$*$] & Oct 5 (6th UT) & Oct 6 (7th UT) & Oct 7 (8th UT)\\ 
    19:00 - 19:30 & A, {\bf B}, C, D, {\bf F} & A, {\bf B}, C, D,{\bf F} & A, {\bf B}, D, {\bf F} \\ 
     19:30 - 20:00 & {\bf A}, C, D, {\bf E}, F & {\bf A}, C, D, {\bf E}, F & {\bf A}, C, D, {\bf E}, F \\ 
      20:00 - 20:30 & C, {\bf D}, F, {\bf G} & {\bf D}, F, {\bf G} & {\bf D}, F, {\bf G} \\
    \hline
    \end{tabular}}\label{tab:tbl2}
\begin{tabnote}
\footnotemark[$*$] Each slot corresponds to 5:00-5:30 UT, 5:30-6:00 UT,  6:00-6:30 UT, respectively. \\ 
\end{tabnote}
\end{table}

\subsection{Automatic Detection}\label{ssec:3_2}
A new meteor detection software, specifically designed for StarCam, was developed by one of our Co-Is (T.U.) after the conclusion of the eye-ball campaign. To evaluate its performance, we decided to test the software on the recorded video data and compare its results with the detections made by the volunteers.

There are many challenges for meteor detection software when processing data from our Starcam. Since the camera's primary purpose is public outreach, its field of view includes the ground landscape. This means that telescopes and passing cars in the scene can cause false detections. While movements from domes are removed by masking, the laser guide star for adaptive optics is particularly prone to being misidentified as a meteor due to its straight structure and frequent changes in brightness. Car headlights are also a problem, as they often illuminate the entire frame, creating a large amount of noise. Additionally, fog near the Maunakea summit, when illuminated by moonlight, can act as a moving light source, leading to false detections. Parameter tuning and filtering to remove these artifacts became a major focus during the software's development. We note that, fortunately, the events that confuses the software did not happen during the evaluation period.

In addition, using YouTube as a live-streaming data source presents its own set of challenges. Real-time detection requires the video to be processed at a speed that keeps pace with the streaming rate. YouTube Live streams are also occasionally unstable, so a system is needed to constantly monitor for disconnections and reconnections. YouTube itself sometimes changes its streaming specifications without announcement, which can lead to unexpected problems. The video data is in MPEG format, so its noise characteristics aren't ideal, with issues like block noise. Furthermore, events that generate a large number of false signals, such as car headlights or lightning, can sometimes overload the computer and cause issues.

The following describes the fundamental principles of the motion detection software. The software was developed in a Python 3 environment, and the detection workflow is as follows.

First, 1920$\times$1080 pixel color frames are acquired from YouTube at 1/30-second intervals and then converted to grayscale. Unwanted areas such as dome structures and the ground are masked out for each frame. Next, contour detection is performed using the \texttt{findContours} function in OpenCV library\footnote{\url{https://opencv.org/}}. While the detected contours contain a significant amount of noise, most noise exhibits random, pixel-level behavior, so objects with a small contour size are excluded. For a genuine meteor, the detected contour is elongated and exhibits a characteristic signal of moving continuously over time along the meteor's path. Therefore, we find a circumscribed ellipse for each detected contour and focus only on objects with an eccentricity greater than 3. A case where such a signal moves continuously in a directional manner over a certain time unit is considered a meteor candidate. This frame-by-frame and inter-frame signal extraction and evaluation creates a list of meteor candidates. These candidates are then further processed to remove false positives like lasers and satellites using additional criteria such as velocity and color. The final list of meteor candidates, including the approximate time, start and end XY coordinates, velocity, and a brightness count, is then saved to a database.

We converted these XY coordinates into RA/DEC data using the calculated World Coordinate System (WCS) information and then measured the shortest distance between the predicted radiant position and the meteor trajectory by tracing the path backward. The WCS for a representative (stacked \& cleaned in a certain time) image was determined using astrometry.net\footnote{\url{https://astrometry.net/}}, and the meteor's position was then pinpointed using a corrected WCS calculated for the specific time of the meteor's appearance. The Zenith Attraction effect was taken into account when calculating the distance to the radiant.

The counts were then aggregated, similar to the visual observation campaign, as a 10-minute count for both Arid shower meteors and the other (mostly sporadic: thus we simply refer these non-Arid meteor as "sporadic") meteors.

\section{Result}\label{sec:4}
Here we will discuss the results of the evaluation from the previous section. We will first present the results from the visual Eye-ball evaluation, followed by a discussion of the re-evaluation using the automatic detection software that was performed later. We note that the solar longitude at 19:30 HST on October 6 in Hawai'i is 193.88 degrees.

\subsection{Eye-ball Detection}\label{ssec:4_1}
\begin{figure}
 \begin{center}
\includegraphics[width=0.49\textwidth]{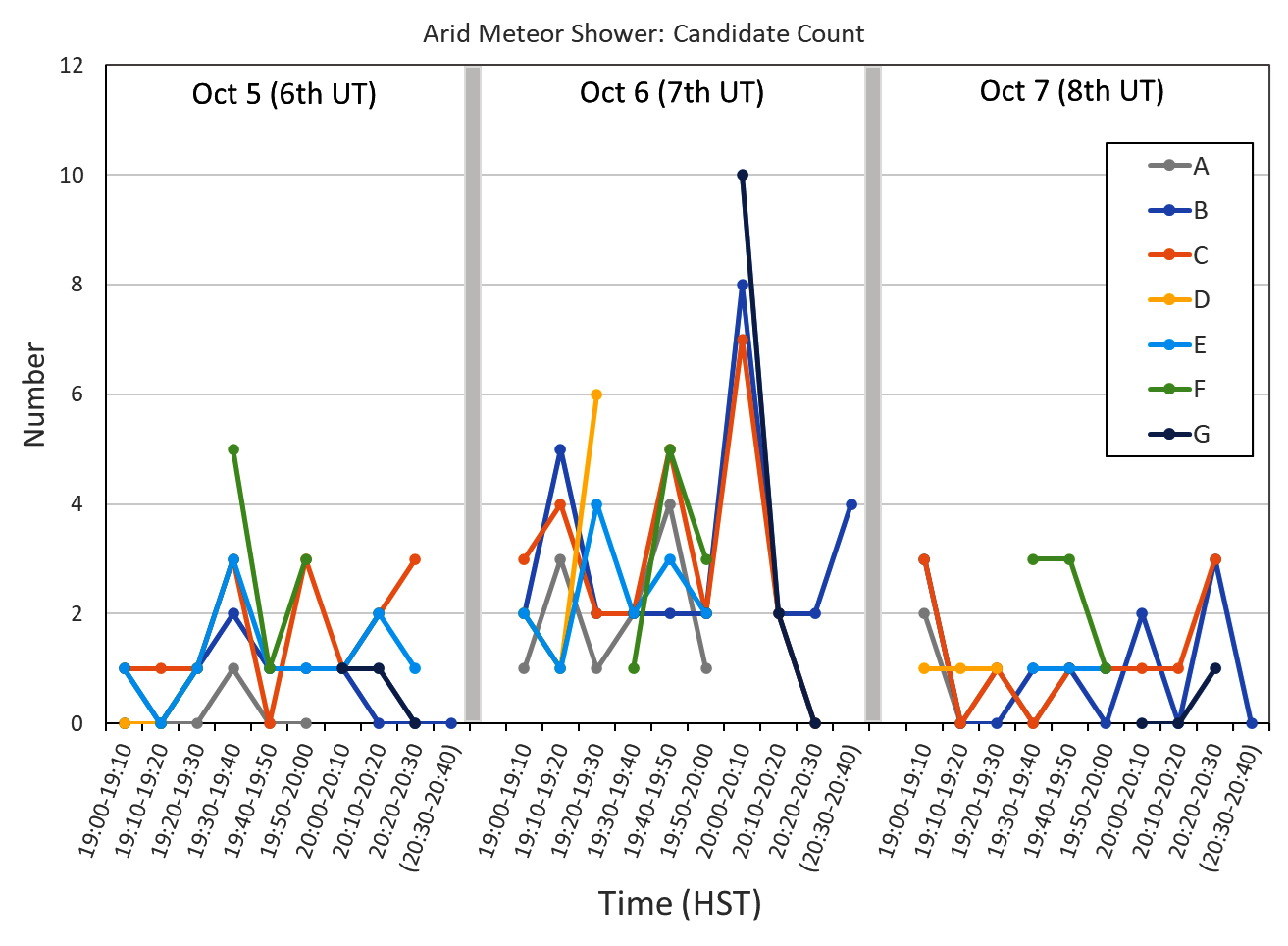}
\vspace{0mm} 
\includegraphics[width=0.49\textwidth]{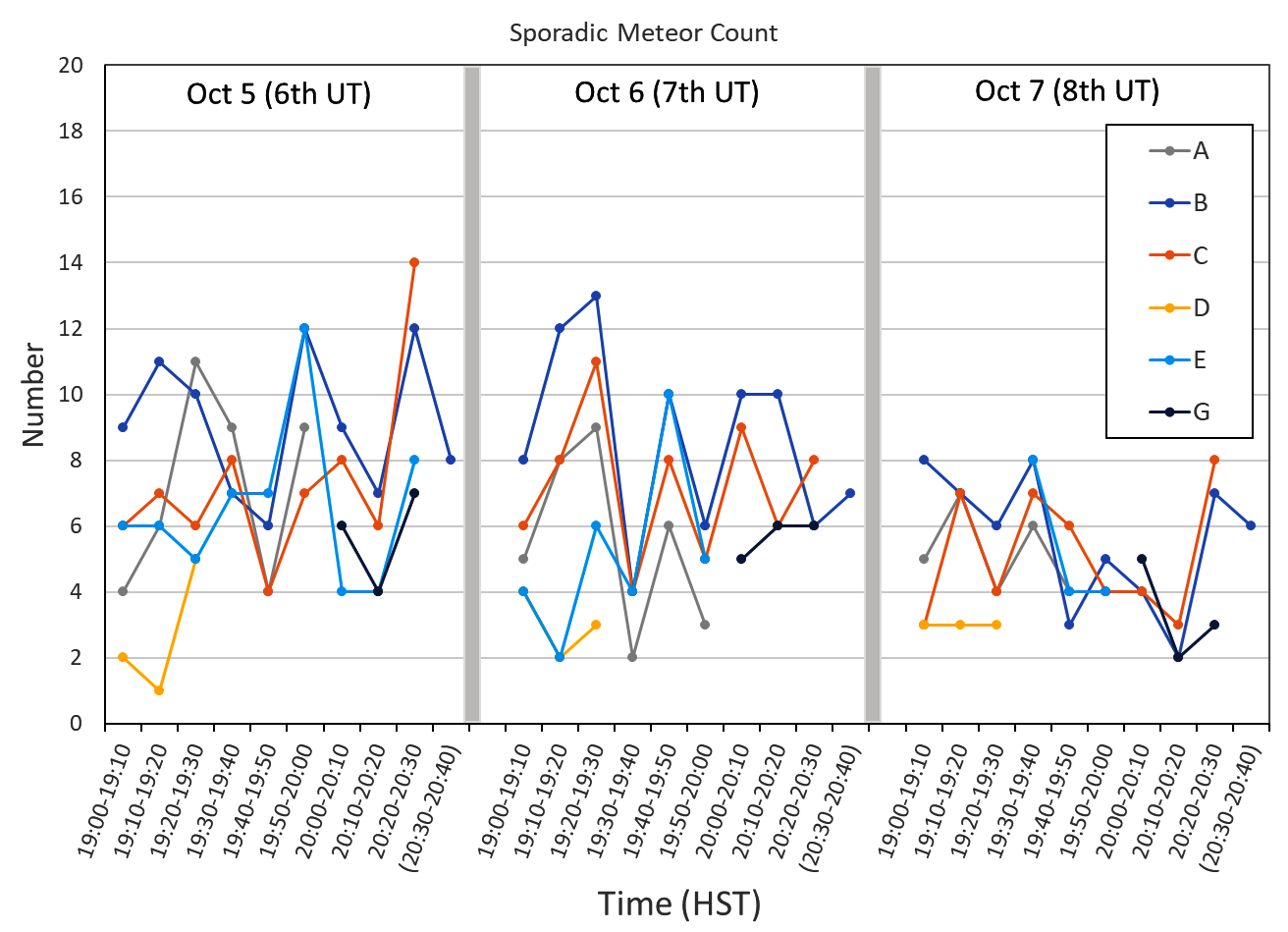}
 \end{center}
\caption{The Arids (Top) and sporadic (Bottom) meteor counts from seven volunteer evaluators. The horizontal axis shows time and date in Hawai'i Standard Time (HST) (= UT - 10 hours). Each volunteer was randomly assigned to evaluate a 30-minute window each day, but many went beyond their assignments. Note that sporadic meteor counts for evaluator F are not available.
{Alt text: This figures are about the result of the eye-ball detection of meteors by volunteers. The individual counts are shown here.} 
}\label{fig:fig2}
\end{figure}

Let's first discuss the results of the visual evaluation by the volunteers. Figure~\ref{fig:fig2} shows the summarized results from each evaluator in a graph. Note that the results are shown for the Arid meteor candidates (Top) and all the other meteors (Bottom). For the latter we put the label "Sporadic" despite that minor fraction ($\sim5\%$) of them belongs to other meteor shower activity\footnote{This is esimated from the result of the auto-detection and evaluation of the data between Oct 1-6, 2024 (clear dark nights)}.

It is evident that there is considerable scatter in the individual detections for both candidate Arids and sporadic meteors. This is likely due to the fact that many meteors were faint and had short trajectories. When focusing on the center of the field of view, evaluators tended to miss meteors that appeared near the edges. Variations in the visual assessment environment, including display size and room brightness, may also influence these results. The slightly lower average counts on the third day can be attributed to the thin cirrus clouds at low altitude, which obscured part of the field of view.

\begin{figure*}
 \begin{center}
\includegraphics[width=0.8\textwidth]{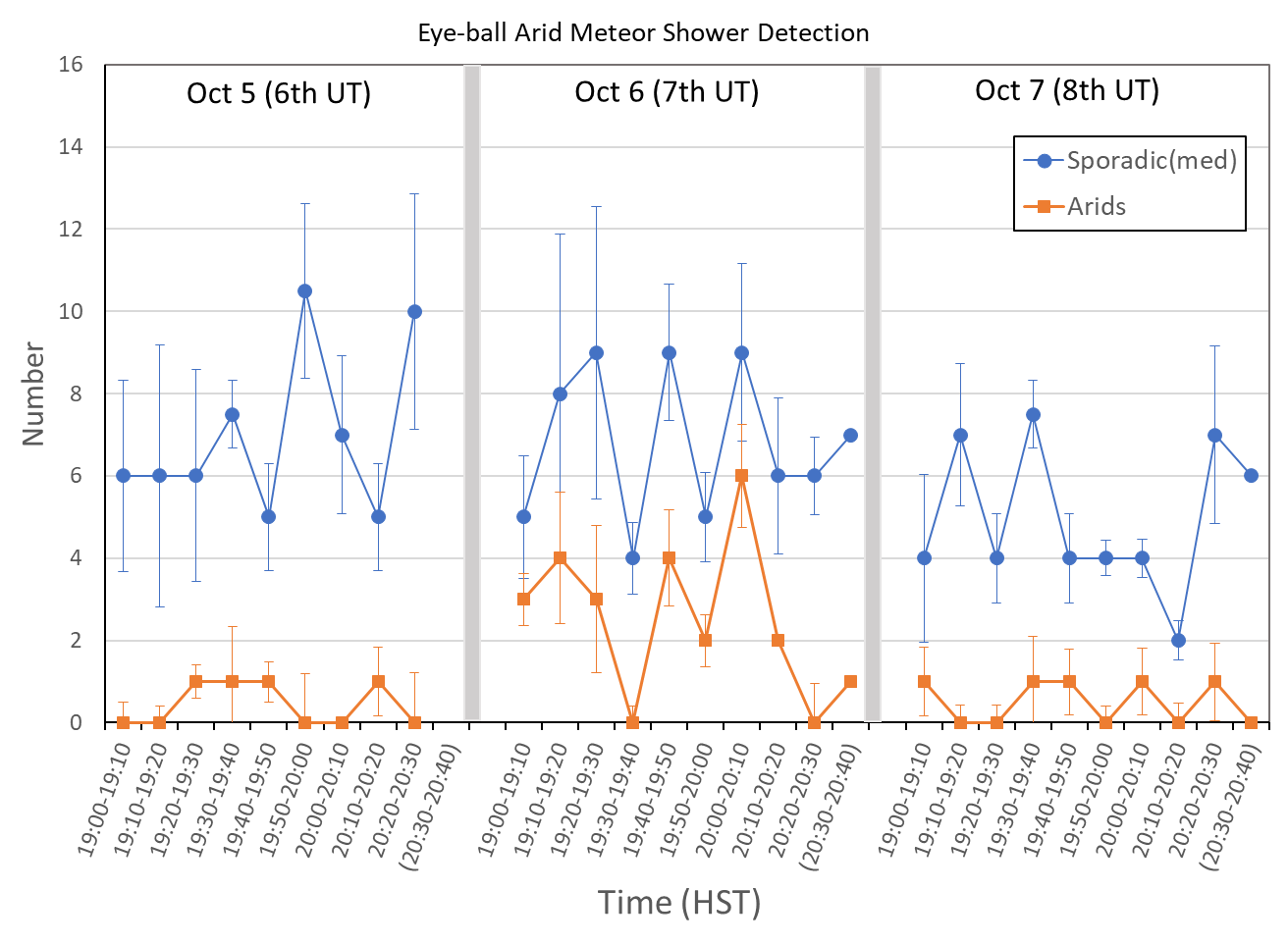}
 \end{center}
\caption{Final results of the visual Arids detection campaign by volunteers. The plot shows a higher number of Arid candidates on the second day (Oct 7 UT), which was the predicted peak date for the shower. A comparison with sporadic meteors also clearly shows a relative increase in detections on this day. Although one evaluator counted meteors during the 20:30-20:40 HST time slot, this data was not included in the final evaluation.
{Alt text: This figures are about the final result (after check) of the eye-ball detection of meteors by volunteers.} 
}\label{fig:fig3}
\end{figure*}

All candidate Arids were re-evaluated by two of the authors (I.T. and M.S.) using the recorded timestamps. As a result, the number of meteors that were removed from the Arids candidate list was 9 on October 5, 13 on October 6, and 6 on October 7 (HST). This suggests that approximately 10 sporadic meteors per 90-minute interval each day have trajectories very similar to the Arids. After this cross-checking correction, the final number of Arids was determined by the remaining counts (with error bars estimated based on the individual volunteer counts). For the sporadic meteors, the final result was determined by taking the median and variance of the volunteer counts. Figure~\ref{fig:fig3} shows the final results of the visual Arids detections.

It is clear from this figure that the number of the Arid candidates on the second day is higher than on the preceding and following days. In contrast, no such trend is observed in the sporadic meteor counts. The total number of the Arid (and sporadic) meteors detected over a 90-minute period on October 5, 6, and 7 (HST) was 4 (63), 24 (61), and 5 (44), respectively. Assuming the 4-5 Arid candidates observed on the other days are representative of sporadic meteors randomly originating from the Arids radiant direction, the detection of 24 meteors from the radiant over a 90-minute period on October 7 (UT) indicates an event of 5--6 times the expected background rate. Although dealing with small number statistics, this excess activity corresponds to a highly significant (potentially $\sim9 \sigma$) event. It is therefore highly unlikely that this result was caused by a random occurrence of sporadic meteors. We thus conclude that the Arid meteor shower was successfully detected in Hawai'i.

\subsection{Automatic Meteor Detection Result}\label{ssec:4_2}

\begin{figure}

\begin{center}
\includegraphics[width=1.0\linewidth]{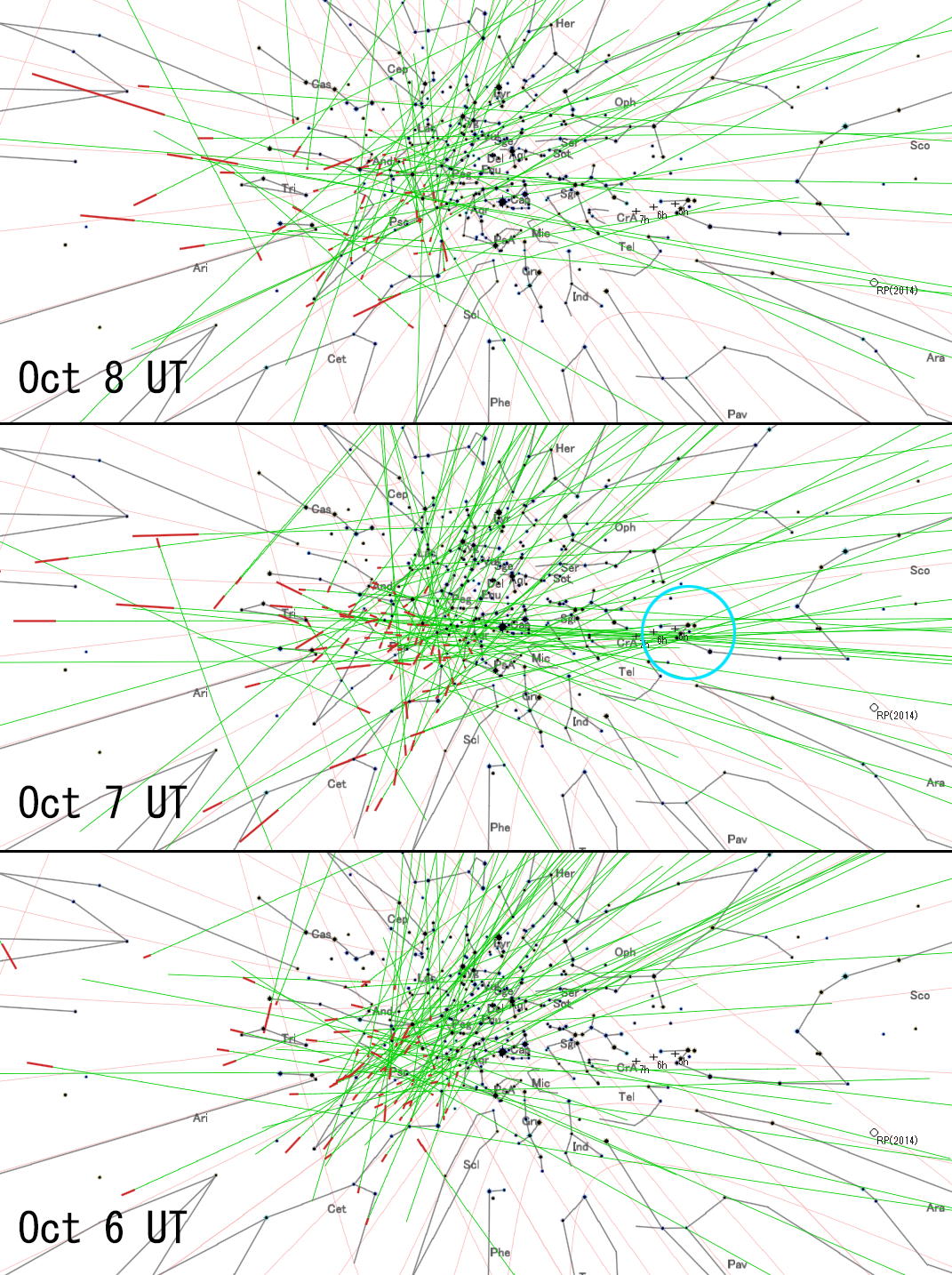}
\end{center}
\caption{This diagram shows the great circles (long lines in green) extending the trajectories of detected meteors (short line segments in red), plotted in horizontal coordinates using a gnomonic projection. From bottom to top, each plot shows the tragectories of meteors detected between 19:00 and 20:30 HST on October 6, 7, and 8, respectively. 
{Alt text: This figure shows the distribution of the grat circle defined by each detected meteor. On 7th UT we can see a concentration of the crossing points of tragectories.} 
}\label{fig:fig4m}

\end{figure}

We then discuss the results of the Arids detection evaluation using the automated detection \& astrometry software. Using the method described in Chapter~\ref{sec:3}, we first show how each great circle defined by the tragectory of each meteor distribute on the sky. The result is in figure~\ref{fig:fig4m}. From bottom to top, each plot shows the tragectories of meteors detected between 19:00 and 20:30 HST on October 6, 7, and 8, respectively. A concentration of trajectories characteristic of a meteor shower is visible on the 7th, indicated by a open circle in the map.

\begin{figure}
\begin{center}
\includegraphics[width=1.0\linewidth]{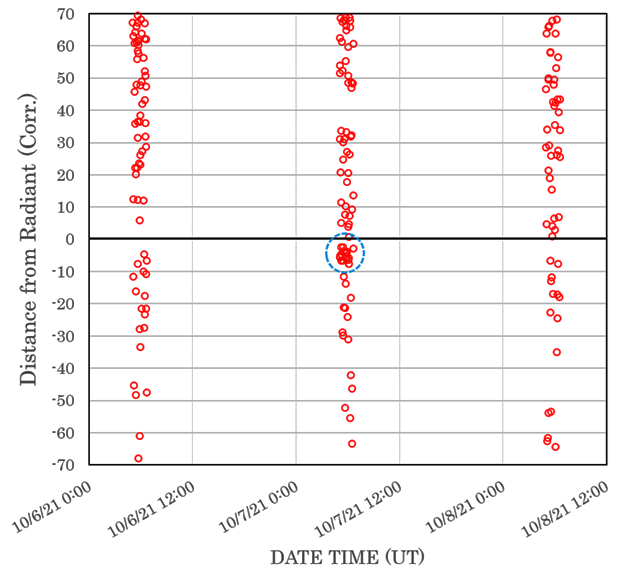}
\end{center}
\caption{Relationship between meteor trajectories and the shortest distance to the provisional radiant point. The horizontal axis represents time in UT. The zenith attraction effect was applied to the provisional radiant's position, and the shortest distance vector is considered positive when it lies to the north. A clustering of the Arid meteor candidate near zero for 7th (circled in dashed line) is evident.
{Alt text: This figure shows the distribution of minimum distance between individual meteor trajectory and tentatively-assigned Arid radiant point.} 
}\label{fig:fig4}
\end{figure}

We then traced the trajectories of the detected meteors backward to a tentatively set radiant point at ($\alpha, \delta$) = (256, -48). We calculated the shortest distance between the extended meteor trajectory and the radiant (after the zenith attraction effect is considered) to see if the meteors clustered. Figure~\ref{fig:fig4} shows the result.

Looking at the vicinity of distance zero in figure~\ref{fig:fig4}, a tight clustering of data points near zero is clearly visible only on the second day (centered at -5.3 $\pm$ 1.7 degrees). The lack of similar data clustering on the preceding and following days strongly suggests that there was unique activity near this radiant on October 7 UT. We believe the clustering does not center on zero due to the astrometric error, because the Arid radiant is in the southwestern sky, while our camera is pointed almost directly east. This large separation angle tends to magnify small measurement errors and systematic astrometry errors.

We evaluated the root mean square (rms) of this tight clustering after a 3-$\sigma$ clipping, which was found to be 1.6 degrees. Meteors falling within a range of $\pm5$ degrees (three times the rms) from the center of this clustering were designated as the Arid candidates.

\begin{figure}
\begin{center}
\includegraphics[width=1.0\linewidth]{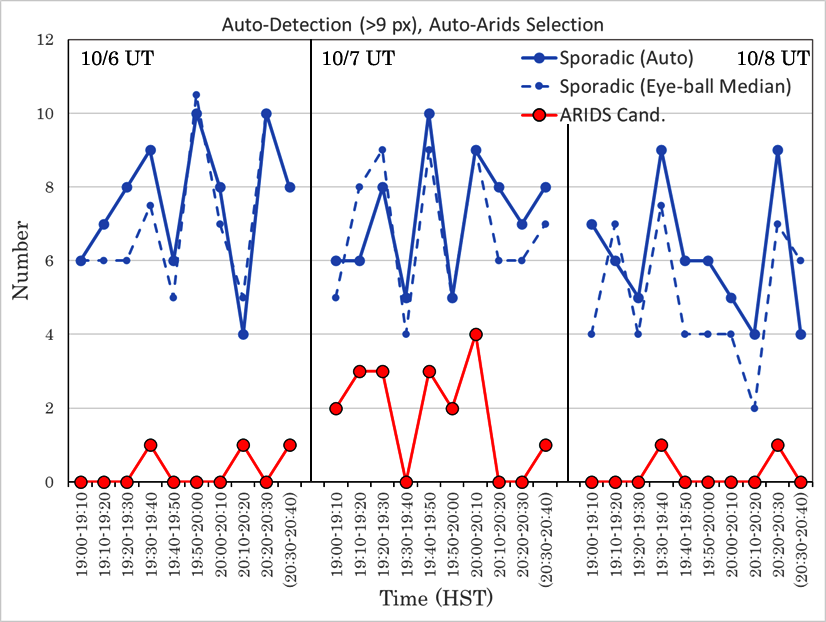}
\end{center}
\caption{This graph shows the 10-minute counts for the Arid candidates (red circles, lower counts) and sporadic meteors (blue line, higher counts) from the automated detection. The median count from the eye-ball detection is also shown for sporadic meteors as a dashed line for comparison. Note that while the dates on the graph are in UT, the times are listed in HST.
{Alt text: This figure shows the result of the automated meteor detection.} 
}\label{fig:fig5}
\end{figure}

Figure~\ref{fig:fig5} shows the 10-minute counts for both Arid and sporadic meteor candidates from the automated detection. It is clear that the number of Arid candidates, shown by the red line, significantly increases on the night of October 7 UT. In contrast, there is no significant difference in the number of sporadic meteors throughout the evaluation period. The slight decrease in the number of sporadic meteors on the 8th UT is likely due to the thin cirrus clouds visible in part of the field of view, consistent with the visual observations.

Based on the automated detection counts, we can draw similar conclusions to the visual observations. The number of the Arid candidates on the second day is significantly higher compared to the days before and after, while no such trend is observed in the sporadic meteor counts. For each 90-minute period on October 5, 6, and 7 (HST), the total number of detected Arids (and sporadics) are 3 (76), 18 (72), and 2 (61), respectively. If we assume that the counts from the preceding and following days reflect the random probability of sporadic meteors appearing from the Arids radiant, the 18 meteors observed on October 7 UT represent a $\sim 6\times$ increase over a random event ($\sim 9\sigma$ if pure Poisson statistics is assumed). It is therefore highly unlikely that this was a random occurrence. Thus, the automated detection confirms that the Arid meteor shower was captured in Hawai'i.

\begin{figure*}
\begin{center}
\includegraphics[width=0.85\linewidth]{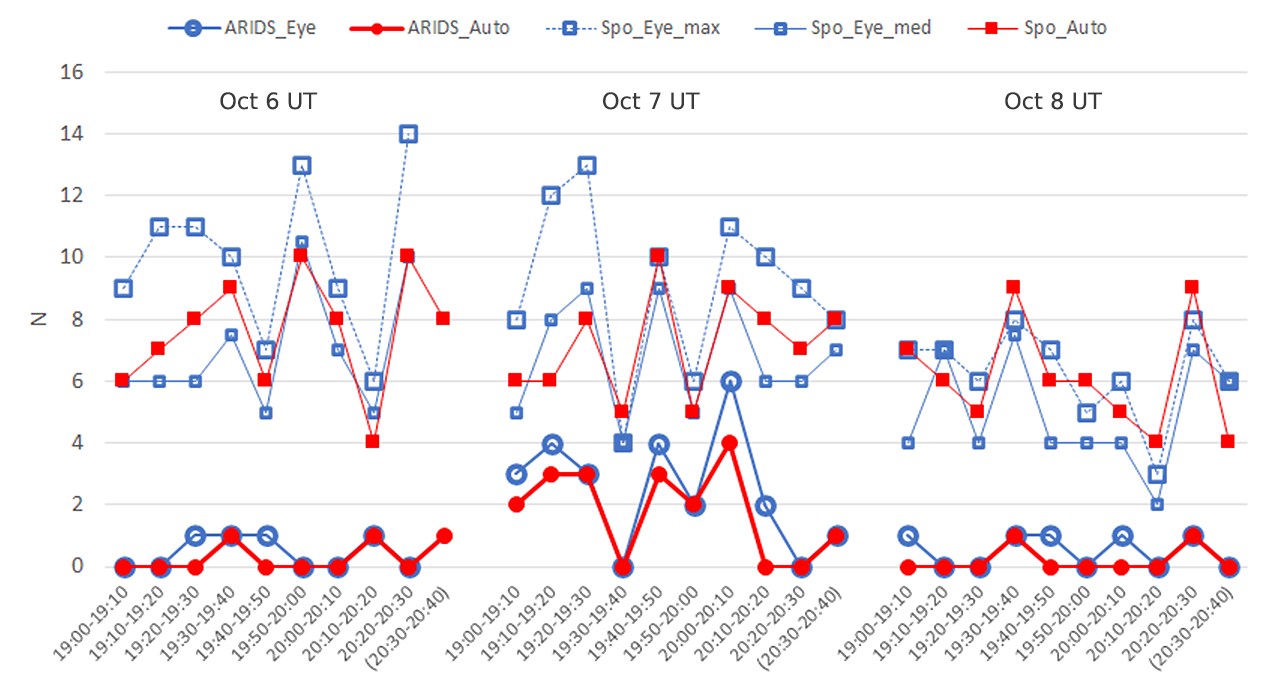}
\end{center}
\caption{This graph compares the automated detection results with those from human eye-ball detection. Solid circles and open circles represent the 10-minute counts of the Arid candidates from the automated and eye-ball detections, respectively. Solid boxes and open boxes represent the sporadic meteor counts from the automated and eye-ball detections. For the open boxes, the marks connected with solid lines show the median count from all evaluators, while the those connected with dashed lines indicates the maximum count. Note that while the dates on the graph are in UT, the times on the horizontal axis are in HST (= UT - 10 hours).
{Alt text: This figure comapres the count results by auto-detection and human detection of meteors.} 
}\label{fig:fig6}
\end{figure*}

Figure~\ref{fig:fig6} shows a comparison of the counting results with those from the visual observations. The overall agreement can be described as fair. For the Arid candidates, there is a tendency for human observers to count slightly more, but it is important to note that the automated detection is not necessarily the more accurate result, as it could miss meteors below a certain length or noisy condition.

For sporadic meteors, the graph shows both the maximum and median counts from the multiple human evaluators, while the automated detection results are often found between these two values. Both human and automated methods tend to miss small meteors, so the general similarity in their trends can be seen as a reflection of the reliability of the counting data.

\subsection{Nature of the Detected Arid Meteors}\label{ssec:4_3}

The Arid meteor shower was predicted to consist of faint, slow-moving meteors, attributed to the low entry velocity of its Jupiter-family comet parent body (e.g., \cite{2020JIMO...48...29V}). The Arid meteors we detected also gave the impression of being relatively faint and slow. Although they were slow, there were several comparatively bright meteors that left short trails or wakes. Moreover, meteors appearing in the evening are generally slower due to the effect by Earth's motion.

At the time of evaluation, our automated detection software had not been absolutely calibrated for magnitude, but the total integrated counts along the meteor's path, after sky subtraction, was only calculated. In addition, the path length (in pixels) and the duration were also measured. From these results, we attempted to gain some insight into the characteristics of the detected Arid meteors.

\begin{figure}
\begin{center}
\includegraphics[width=1.0\linewidth]{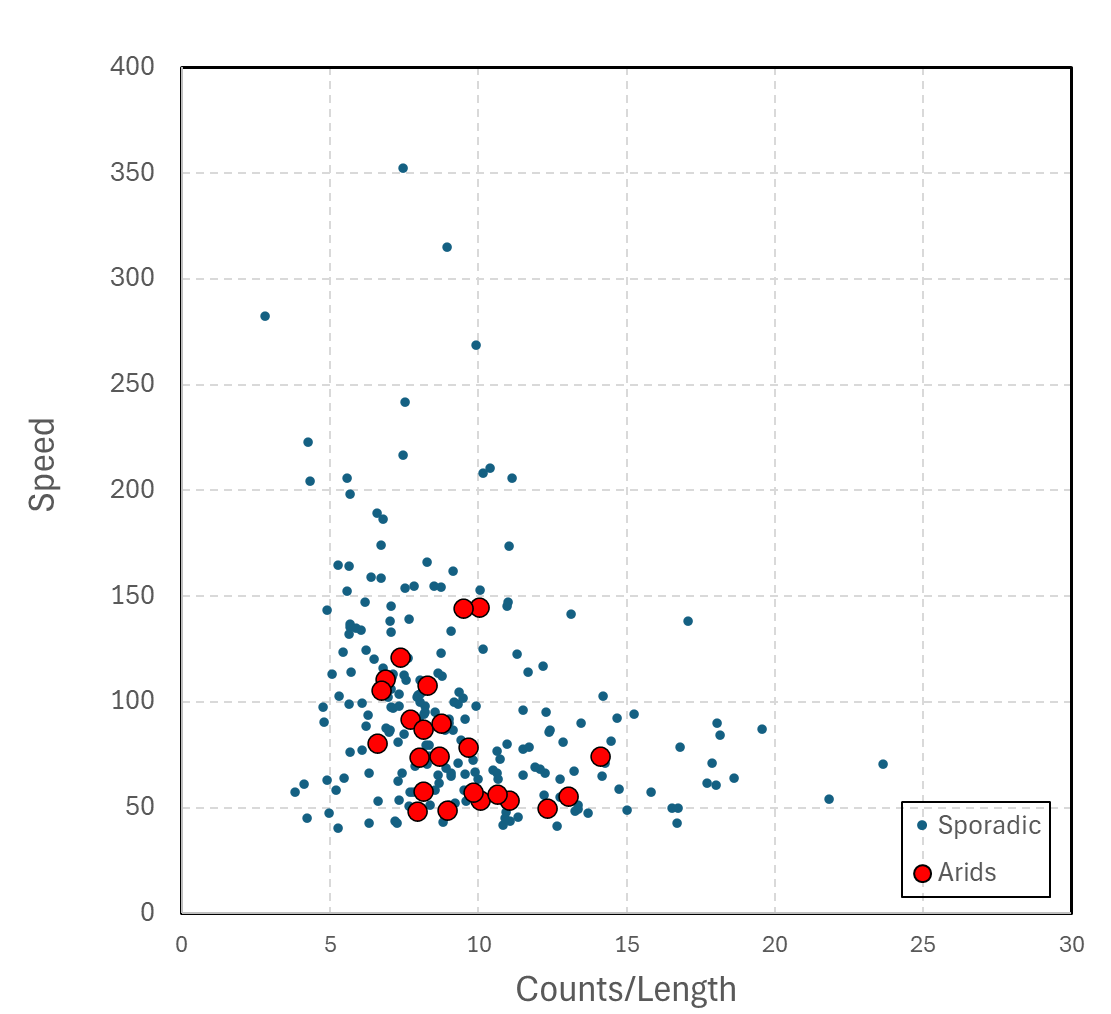}
\end{center}
\caption{The relationship between the average brightness of meteors and their velocity (in pixels/sec) during the evaluation period. Sporadic meteors are represented by small blue circles, while the Arid meteors are shown as large red circles. The horizontal axis indicates average brightness, with brighter meteors to the right.
{Alt text: This figure is the comparison of the Arids and sporadic meteors in the plane of average meteor brightness and the speed.} 
}\label{fig:fig7}
\end{figure}

Figure~\ref{fig:fig7} shows the relationship between a meteor's velocity (in pixels/sec) and its average brightness. The average brightness is calculated by dividing the total integrated count over the entire path by the path length. On the horizontal axis, brighter meteors are to the right. The absence of data points to the left of 5 Counts/Length represents the detection limit. On the vertical axis, it can be seen that velocities below approximately 40 pixels/sec are filtered out.

When comparing the Arids with sporadic meteors, the Arids appear to trace the same distribution as the sporadic meteors. In other words, it cannot be said that there is a significantly higher number of fainter and slower meteors compared to the sporadic meteors typically observed in the evening, many of which appear slow due to the effect of Earth's motion.

However, observational effects must be considered. Our camera's field of view is pointed at the eastern sky, more than 90 degrees away from the Arid radiant in the low, south-southwestern sky. The Arid meteors observed in this field of view are seen from a side-on perspective as they enter the Earth's atmosphere, which means their velocity tends to be observed at its maximum. In contrast, sporadic meteors arrive randomly, and their observed velocity is only the tangential component, which creates a strong effect of appearing to be slower. Therefore, we cannot conclude that the actual entry velocities are the same just because the observed velocity distributions are similar. Rather, if the velocity distributions are the same, it would suggest that the \textit{actual} average entry velocity of the Aris meteors is, in fact, slower than that of sporadic meteors.

How should we interpret the nearly threefold range in the observed velocity of Arids? This is largely due to the wide field of view and the fact that the camera is observing at low altitudes. Meteors at low altitudes is actually rather distant. When a meteor occured at an altitude of 90 km in vertical distance is observed at the top edge of our field of view, its distance from the observation point is about 140 km. However, for a meteor at the bottom third of the field of view, the distance can range from 350 km to over 600 km. Since more distant meteors appear slower, this can account for a three-fold velocity difference between the top and bottom of the field of view. Indeed, the three fastest Arid meteors observed were all located near the top edge of the field. The same applies to sporadic meteors, but the tail of high-velocity meteors (>150 pixels/sec) visible only in the sporadic distribution suggests that Arid candidates do not have velocities comparable to fast sporadic meteors.

Regarding the average brightness distribution, it appears to be similar to that of sporadic meteors. We cannot say that the Arids are richer in faint meteors compared to sporadics. Some meteors were actually fairly bright, leaving short trails. While further investigation into the brightness distribution is warranted, we must acknowledge the inherent limitations of a single-point observation.

\section{Discussion}\label{sec:5}
We now continue our discussion of the detection results. The appearance of the Arid meteor shower was reported by \citet{2021eMetN...6..534J} to have peaked at a solar longitude of 193.68 $\pm$ 0.17 degrees, specifically at 00:41 UT on October 7. A key characteristic of this shower is its relatively slow geocentric speed of 10.5 $\pm$ 0.3~km s$^{-1}$. Furthermore, radio observations by \citet{2023PSJ.....4..165J}
reported activity lasting over seven hours, centered around 01:00 UT. \citet{2021eMetN...6..578O} also reported strong activity between 0 - 1~h UT. This allows us to conclude that there was activity peaking around 01:00 UT, which was close to the time predicted by Sato, Ye, Maslov, and Vaubaillon\footnote{See Ye et al. 2021, ATel \#14947, \url{https://astronomerstelegram.org}}.

In contrast, our observations were conducted from 05:00-06:30 UT on October 7 (solar longitude = 193.9), approximately 4-5 hours after the reported peak activity. Based on the radio activity evaluation in figure~1 of \citet{2023PSJ.....4..165J}, the activity during our observation window is suggested to have decayed to about one-tenth of its peak level. The activity we found for the Arids, with a flux of about half that of all other meteors (mostly sporadic meteors), is consistent with this final stage of activity.

If a similar correlation exists between radio and visual observations, a tenfold increase in activity at the peak time would mean the  Arid shower would have had a flux $2.5\sim4$ times greater than the activity of sporadic meteors in early evening, making it a fairly active shower (see sections~\ref{ssec:4_2} \& \ref{ssec:4_3}: our observed Arid meteor shower flux is 25 to 39 \% of sporadic meteors). To put this into perspective, we can compare it to the Geminid meteor shower. We referred to our Starcam data from December 13, 2023 (new moon), the day of the Geminid peak. During the period when the Geminid radiant was at an altitude of 20-40 degrees (21:00-22:30 HST), we counted 281 Geminid meteors while 161 other meteors. This may suggest that if the activity of the Arids at its peak was observed by our camera, and if the activity levels of radio and visible meteors are proportional, the Arids could have been twice as active as the Geminids.

Of course, the relationship between radio and visible activity is likely not that simple, so this conclusion might be an overestimation. However, Pablo Vera (U. de La Serena) reported visually counting 35 meteors per hour through breaks in the clouds during the peak time at the Observatorio El Sauce in Chile\footnote{\url{https://www.imcce.fr/recherche/campagnes-observations/meteors/2021arids}}). This suggests that the activity was indeed quite distinct.

\begin{figure}
\begin{center}
\includegraphics[width=0.75\linewidth]{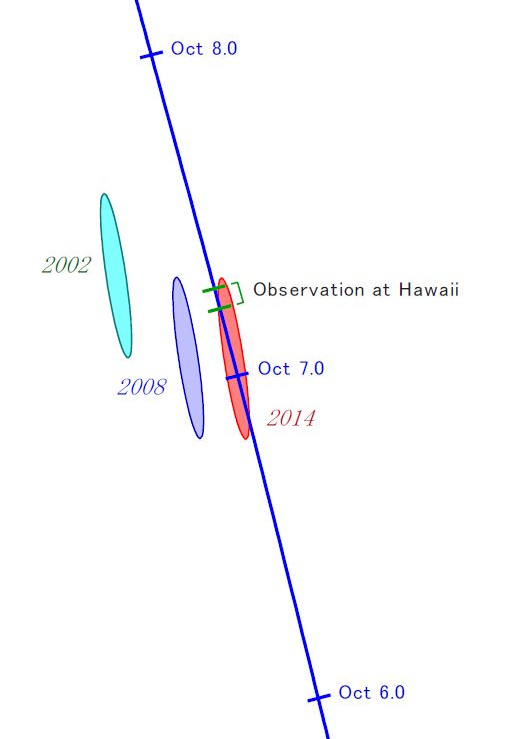}
\end{center}
\caption{The result of a new, detailed simulation of Comet 15P/Finlay's dust trail that reproduce our observations. Dates along the Earth's orbit are in UT. The observation window from Hawai'i is highlighted in green.
{Alt text: This figure shows the result of the new dust-trail simulation for 15P/Finlay that explain our observation. } 
}\label{fig:fig8}
\end{figure}

To investigate the origin of the meteoroids we observed, our Co-I (M.S.) ran a new simulation of the dust trails from 15P/Finlay using a method similar to \citet{2005PASJ...57L..45W}. The results are shown in Table~\ref{tab:tbl3}. Column (1) lists the ejection year of the trail we tracked. Columns (2)–(4) are the date, time, and solar longitude of the closest encounter for each trail. Columns (5)–(7) indicate difference in heliocentric distance in the ecliptic plane, ejection velocity ("+" mean the direction of the comet’s motion), and the degree of dust trail extension (dimensionless) for the trail that resulted in the calculated encounter. Columns (8) and (9) show the position of the radiant, and Column (10) is the impact velocity with Earth.

Our calculation suggest that our detections can be explained by dust trails released in either 2008 or 2014 (see figure~\ref{fig:fig8}, but both scenarios require a considerably high ejection velocity. High-velocity dust ejection would be expected only if the parent comet were extremely active during its return, yet observations of 15P/Finlay's 2008 return suggest its activity was quite subdued\footnote{See Yoshida, S.  \url{http://www.aerith.net/comet/catalog/0015P/2008.html}}. Therefore, it is doubtful that such high-speed dust ejection occurred.

\begin{table*}[htbp]

\caption{Predictions from Dust Trails} \label{tab:tbl3}
\centering
\begin{tabular}{ccccccccccc}
\hline
Ejection & \multicolumn{3}{c}{Expected peak time} & $\Delta r$ & Ejection Velocity & fM & \multicolumn{2}{c}{Expected position} & Vg \\
year & Date (UT) & Time & LS (2000.0) & (AU) & (m s$^{-1}$) &  & $\alpha$ & $\delta$ & (km s$^{-1}$) \\
(1)  & (2)       & (3)   & (4)         & (5)  & (6)   & (7) & (8)      & (9)      & (10) \\
\hline
\hline
2002 & 2021/10/07.38 & 09:05 & 194.025    & -0.0048    & +25.39 & 0.38 & 255.92 & -47.63 & 10.78 \\
2008 & 2021/10/07.08 & 02:01 & 193.734    & -0.0022    & +34.98 & 0.50 & 255.65 & -48.30 & 10.76 \\
2014 & 2021/10/07.05 & 01:12 & 193.701    & +0.000028  & +67.28 & 0.97 & 255.68 & -48.35 & 10.74 \\
\hline
\end{tabular}

\end{table*}

Conversely, if the dust originated from the 2014 return, the required ejection velocity would be an exceptionally high 68~m s$^{-1}$. However, this return showed multiple large outbursts (e.g., \cite{2016AJ....152..169I}), which would likely have produced a significant amount of high-velocity dust. Actually in the case of Herculid meteor shower in 2022, relatively large meteoroids ejected at a large negative velocity of approximately -27~m s$^{-1}$ were observed \citep{sato2025sp}.  Our calculations show that the 2014 dust trail from P/15 Finlay had a minimum approach distance of just 0.000028 AU, which is effectively a collision with the Earth, and therefore a very strong dust flux would be expected at its peak. Furthermore, the comet's small orbital inclination (6.8 degrees) suggests that even a young dust trail could produce long-duration meteor shower activity, which explains why we observed a significant level of activity during our observation window.

Based on these points, we conclude that the meteors we detected in Hawai'i most likely originated from high-velocity dust ejected during the 2014 outburst.

\section{Conclusion}\label{sec:6}
We report the successful detection of the 2021 Arid meteor shower using the Subaru-Asahi StarCam on Maunakea, Hawai'i. The detection campaign involved two methods: a volunteer-led visual inspection and a subsequent re-evaluation with a newly developed automated detection software.

Our observation occurred approximately 4-5 hours after the predicted peak. The number of meteors near the predicted radiant on October 7 UT was more than six times higher (a $\sim9\sigma$ event) than on the days before and after by both methods, suggesting that we successfully captured the receding tail of the activity. This is a quite unique observation achieved from the northern hemisphere. 

Our observation is consistent with radio observation data that suggested activity had decayed to about one-tenth of its peak level by that time. We further speculates that if the activity levels of radio and visible meteors are proportional, the Arid meteor shower at its peak could have been twice as active as the Geminid meteor shower. The detected meteors were generally characterized as faint and slow, though several brighter ones with wakes were also observed.

Regarding the origin of the meteoroid dust, a simulation of the dust trails from the parent body, Comet 15P/Finlay, was performed. This simulation indicated that the detections could be explained by dust trails from either the comet's 2008 or 2014 returns. However, the 2008 return was exceptionally quiet, making the high dust ejection velocities required for that scenario questionable. In contrast, the 2014 return was marked by multiple large outbursts, which would have produced the high-velocity dust needed to account for the observations. Based on these findings, the paper concludes that the meteors detected in Hawai'i most likely originated from the high-velocity dust ejected during the 2014 outburst of Comet 15P/Finlay.

\begin{ack}
First and foremost, we extend our sincere gratitude to our committed volunteers for their participation in the "Detect Arid Meteor Shower" Campaign. We would like to specifically thank Amehare-Chan back, Kage-tan, Cookie, Hiroaki Suzuki, Harutaka Nagahata, Hiroshi Minagawa, and Watashi-Meme. Watashi-Meme's contribution included the preparation of the video data from the live stream during the evaluation period. This study would not have been possible without her dedication. We also thank the viewers of the Astro LIVE by the StarCams. Finally, we thank Dr. J{\'e}r{\'e}mie Vaubaillon, the referee of our paper, for his userful comments and support for publication.
The authors respectfully recognize the profound cultural significance and enduring reverence that the summit of Maunakea holds for the indigenous Hawaiian community. We feel deeply honored to have the opportunity to conduct astronomical observations from this sacred mountain.
\end{ack}



\end{document}